\documentstyle[aps]{revtex}
\input{epsf.sty}
\begin{document}
\begin{titlepage}
\begin{flushright}
LU TP 97/16 \\
NORDITA-97/50 N/P \\
hep-ph/9708232 \\ 
July 1997
\end{flushright}

\vfill

\begin{center}
\begin{bf}
{\Large \bf Matching the Heavy Vector Meson Theory} \\[2cm]
\end{bf}
J. Bijnens$^a$, P. Gosdzinsky$^b$ and P. Talavera$^c$ \\ [1cm]
$^a$Department of Theoretical Physics, University of Lund \\
S\"olvegatan 14A, S-22362 Lund, Sweden. \\[0.5cm]
$^b$NORDITA, Blegdamsvej 17, DK-2100 Copenhagen \O , Denmark. \\ [0.5cm]
$^c$Departament de F{\'\i}sica i Enginyeria Nuclear \\
Universitat Polit{\`e}cnica de Catalunya, E-$08034$ Barcelona, Spain. \\[1cm]

\vfill

\end{center}
\vfill
\begin{abstract}
We show how to obtain a ``heavy'' meson effective
lagrangian for the case where the number of heavy particles is not conserved.
We apply the method in a simple example at tree level and perform
then the reduction for the case of vector mesons in Chiral Perturbation
Theory in a manifestly chiral invariant fashion. Some examples showing that
``heavy'' meson effective theory also works at the one--loop level are
included.
\end{abstract}
\vfill
\end{titlepage}

\setcounter{footnote}{0}
\section{Introduction}
In a well-known paper\cite{IW} Isgur and Wise derived extra consequences of
the heavy mass of a particle in a restricted class of processes. Their method
was then quickly generalized\cite{georgi} and extended for use
in various sectors of Chiral Perturbation Theory \cite{wiseCHPT,jenkins0}.
The general class of processes dealt with in the latter references is with
one ``heavy'' particle going in and moving through the whole process
and going out again. Everything else has low momenta compared to the heavy
particle.
The common point is that the number of ``heavy'' particles remains
constant during the process due to a conserved quantum number. This is not the
case in all processes where we expect this type of expansion to be useful.
A simple but naive example would be the process in which 
intermediate state contains particles of even higher masses. Those can
be integrated out first and present no theoretical problem. We could,
however, study the process $W^+ b\to W^+ b$ in a theory with a light $W$-boson.
Then we would have to consider the charm quark intermediate state. The
charm quark here has a large momentum and the process should be expressible
as a power series in $1/m_b$ in an effective lagrangian where the $b$
quark always takes the large momentum.

A similar problem arises when we try to do Chiral Perturbation Theory for
vector mesons. In reference \cite{ELISABETH}  was conjectured that
a similar formalism should exist there. (See \cite{other} for other
applications of this formalism).
The main obstacle of dealing with this situation is that the methods
used in\cite{georgi}---or e.g. \cite{shift} for the extension
to other spins--- do not take all terms correctly into account in
performing the reduction from the full theory to the ``heavy'' effective one.

In order to estimate the parameters at higher order, in our previous
calculation 
\cite{BGT}
we performed the matching of the relativistic models
with the ``heavy meson'' formulation in a diagrammatic fashion.

The diagram by diagram formulation becomes difficult if we want to do the
determination of all terms resulting from this reduction. In particular,
chiral symmetry relates processes with different number of pions
and it would therefore be useful to have a procedure that fully generates
the correct terms immediately. 
For interactions among pions only when integrating out vector mesons this
can be done
using the equations of motion for the vector
meson field\cite{TONInucl,BBC}. The objective of this paper is to show how
this can be done for the ``heavy meson'' effective theory. The main
complication is that we have to correctly treat the
diagrams with only light intermediate states, one of which then has to carry
high momentum, and at the same time the contributions from the heavy state
to subprocesses involving light momenta only. For the ``light'' and ``heavy''
states in our theory we thus have to keep two possible momenta regimes.
One around the mass-shell of the ``light'' mode and the other around the 
mass-shell of the ``heavy'' mode.
In the usual cases only the momentum regime where all particles were close
to their mass-shell is kept.\footnote{For the pions close to their mass-shell
here means low-momentum.}

Keeping track of several momentum regions of the particles can in
principle be done by introducing several components in the field,
each of which has only
a low momentum:
\begin{equation}
\phi_{\rm full} = e^{-i M v\cdot x} \tilde{\phi_v} +\tilde \phi_0
 + e^{i M v\cdot x} \tilde{\phi_v}^\dagger\,.
\label{simple}
\end{equation}
Notice that effective fields are traditionally normalized
differently.  It corresponds to
$\tilde\phi_v\to\tilde\phi_v/\sqrt{2M}$. This factor should be understood in
what follows.
The component with the large (small) momentum we will refer to as the high
(low) component in order
not to confuse with light (heavy) for the mass.

We then integrate out for the heavy field the component $\tilde\phi_0$ and for
the light field the component $\tilde\phi_v$.
This procedure works at tree level
since there is always only one line carrying the heavy momentum through
the whole diagram. We therefore also only need to keep at most two powers of
the high components. The end result is then an effective theory formulated
in terms of the high component of the heavy field and the low component
of the light field. At any time, terms in the lagrangian which cannot possibly
conserve momentum do not contribute since they vanish after the
integration over all space.

The approach outlined above will obviously work in simple models. In
the chiral models with vertices with any number of fields it becomes
rather cumbersome. We will therefore choose a different method. By introducing
a ``hidden'' symmetry we can introduce an extra spurious degree of freedom.
The ``gauge fixing'' of this symmetry consists in choosing an
equation for the extra degree of freedom
in such a way that the spurious degree of freedom takes care of the far
off-shell components of the fields. 

We will present the main idea in three stages. First we show it in a simple
model where this approach is identical to the one in Eq. (\ref{simple}) but
shows the procedure in a simple way. Then we show the procedure in a model
with an external scalar field coupling to pions in a chirally invariant
way. Finally we present the case for vector mesons interacting with pions.
Here we rederive and generalize the results of \cite{BGT}.

We also show in a few examples that the ``heavy'' meson method works
at the one--loop level and present a short discussion of the relevance of
the width.

\section{A Simple Model}

\label{treesimple}
Consider the following lagrangian 
with a heavy $\phi$ field and a light $\pi$ field:
\begin{equation}
{\cal L}_0 = \frac{1}{2}\partial_\mu \phi\partial^\mu\phi +
\frac{1}{2}\partial_\mu\pi\partial^\mu\pi-\frac{1}{2}M^2\phi^2
-\lambda M \phi \pi \pi
\label{model}
\end{equation}
We will be interested
in the limit, in which the $\phi$ mesons gets very heavy, that is,
$M \rightarrow \infty$. We will consider that $\lambda$ in this
limit is a non-vanishing finite constant.

\begin{figure}
\begin{center}
\leavevmode\epsfxsize=12cm\epsfbox{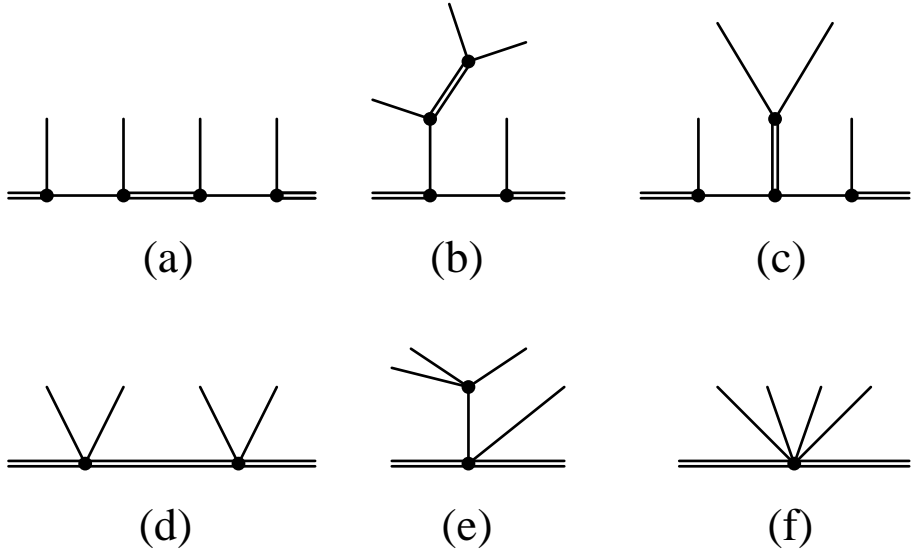}
\end{center}
\caption{\label{figgreen} Six point functions in the Simple Model: (a), 
(b), (c) are in the relativistic formulation and (d), (e) and (f) are their
respective counterparts in the HMT.
Double lines correspond to the ``heavy'' field and
full lines to the ``light'' field.}
\end{figure}

The ``heavy'' meson theory (HMT) now looks at processes of the type
\begin{equation}
\phi \,+\,n\,\pi \longrightarrow \phi\,+\,k\,\pi\,,
\label{processes}
\end{equation}
with the $\phi$ on-shell and the momenta of the pions small compared
to the $\phi$-mass $M$. 
$n\,\pi \longrightarrow \,k\,\pi$ processes are relevant as well, they occur
as subprocesses in (\ref{processes}).
An important point is to notice that
processes such as a decay of
a $ \phi $ into two $\pi$'s, allowed by (\ref{model}), do not lie
within the heavy meson theory. 
 The vertex $\lambda M \phi \pi ^2 $ can, however, generate Green
functions that lie within the HMT.  In figure \ref{figgreen}c, for example,
we have combined four of these vertices to construct a six--point
function that is described within the HMT. This Green function cannot
be generated in the same way in the  heavy meson theory, and we will
see how the HMT-vertices that accounts for the Green function can be
generated. 

The naive limit to the heavy meson theory would be to just write
\begin{equation}
\phi = e^{-i M v\cdot x}\tilde{\phi} + e^{i M v\cdot x} \tilde{\phi}^\dagger
\label{naive}
\end{equation}
and restrict $\tilde{\phi}$ and $\pi$ to momenta much smaller than $Mv$.
Here $v$ is a four-velocity as introduced in \cite{IW},
$v^2 = 1$. For ${\cal L}_0$ this naive procedure would lead to a lagrangian
without interaction terms\footnote{The interaction term always vanishes
in this approximation because of momentum conservation.},
an obviously wrong conclusion.
The method where we also keep the particular far off-shell regions relevant
for the processes of Eq. (\ref{processes}) as in Eq. (\ref{simple})
can be used here and gives the same results as a diagram by diagram matching.

A different way to achieve the same result is introducing first 
two extra symmetries $R_1\times R_2$ with the extra fields $\sigma$ and $\psi$.
The symmetry transformations are with $\alpha_1\in R_1$ and $\alpha_2\in R_2$:
\begin{equation}
\phi\to\phi+\alpha_1,\quad\psi\to\psi-\alpha_1,\quad\pi\to\pi+\alpha_2,
\quad{\rm and}~~~\sigma\to\sigma-\alpha_2\,.
\end{equation}
The lagrangian,
\begin{eqnarray}
{\cal L}_1 &=& \frac{1}{2}\partial_\mu \Phi\partial^\mu\Phi +
\frac{1}{2}\partial_\mu\Pi\partial^\mu\Pi-\frac{1}{2}M^2\Phi^2
-M \lambda\Phi\Pi\Pi
\nonumber\\
\Phi& =& \phi+\psi\nonumber\\
\Pi&=&\pi+\sigma\,,
\label{L1}
\end{eqnarray}
is equivalent to ${\cal L}_0$ and has the extra symmetry. We can simply
return to ${\cal L}_0$ by choosing $\psi=\sigma=0$.
We use the freedom of gauge to have $\psi$ take care of the low part of $\phi$
and $\sigma$ of the high part of $\pi$. We choose:
\begin{eqnarray}
0&=&-\partial^2\psi-M^2\psi-\lambda M \pi\pi-\lambda M 
\sigma\sigma 
\nonumber\\
0&=&-\partial^2\sigma-2\lambda M \phi\pi-2\lambda M \psi\sigma\,.
\label{choice}
\end{eqnarray}
Here it  should be understood that
inside the $ \sigma\sigma $ term only the ``low momentum part'' is taken, 
(i.e. the term behaving like: $2 \tilde{\sigma} \tilde{\sigma}^\dagger $,
see below).
This choice now allows us to make the {\em consistent} set of approximations:
\begin{eqnarray}
\pi &=& \tilde\pi\nonumber\\
\psi&=&\tilde\psi\nonumber\\
\phi &=& e^{-i M v\cdot x}\tilde{\phi} + e^{i M v\cdot x} \tilde{\phi}^\dagger
\nonumber\\
\sigma&=& e^{-i M v\cdot x}\tilde{\sigma} 
      + e^{i M v\cdot x} \tilde{\sigma}^\dagger\,.
\label{consistent}
\end{eqnarray}
This is consistent since the equations of motion are now such that
all terms in the equations that contribute to the processes (\ref{processes})
are correctly taken care of. The problem before was that, if we used
Eq. (\ref{naive}) only, there were ``driving'' terms in the equations
that had the wrong momentum\footnote{The other terms can be neglected since
at tree level we can never have intermediate lines with momenta far away
from either 0 or $Mv$.} that had nowhere to go. The choice of gauge
for the spurious fields of (\ref{choice}) solves this problem.
All the tilded fields in (\ref{consistent}) are ``low momentum fields'', that
is, their Fourier transforms are only non-vanishing for momenta 
$ p \sim 0$. It is this fact that will allow us to perform a consistent
and well defined expansion in $(1/M)$.

The ``heavy'' lagrangian should only depend on the fields
close to their mass--shell.
We therefore integrate out $\tilde \psi$, 
$\tilde \sigma$ and $\tilde \sigma ^\dagger$.
At tree level, this can be done by choosing the gauge 
and replacing the solution in the lagrangian. We have
solved (\ref{choice}) by performing an expansion in $1/M$. Up to
order ${\cal O} ( 1/M^3 ) $, we find:
\begin{eqnarray}
\tilde \psi & =& 
-\frac{\lambda}{M} \tilde \pi ^2 + \frac{\lambda}{M^3} \Box \tilde \pi^2
       -8 \frac{\lambda^3}{M^3} \tilde \pi ^2
      \tilde \phi \tilde \phi^\dagger  \label{sol1OM} \\
\tilde{\sigma}& =& 2\frac{\lambda}{M} \tilde \pi \tilde \phi 
    - 4 i \frac{\lambda}{M^2}
               v \cdot \partial(\tilde \pi \tilde \phi ) +
        \frac{\lambda }{M^3} \left( 2  \Box (\tilde  \pi \tilde \phi ) 
         -8 (v \cdot \partial)^2 (\tilde \pi \tilde \phi) -
         4 \lambda^2 \tilde \phi  \tilde \pi ^3  \right) \nonumber
\end{eqnarray}
To go to the nonrelativistic limit, we introduce the correct
normalization,
\begin{equation}
\bar \phi = {1 \over \sqrt{2 M} } \tilde \phi, \qquad
\bar \phi ^\dagger = {1 \over \sqrt{2 M} } \tilde \phi ^\dagger .
\label{nrPHI}
\end{equation}
Replacing (\ref{consistent}), (\ref{sol1OM}) and (\ref{nrPHI}) 
in (\ref{L1}), and
expanding up to order ${\cal O}(1/M^2)$, we find:
\begin{eqnarray}
{\cal L}_1^E &=&  {1 \over 2} \partial _\mu \tilde \pi ^2 +
{1\over 2 M} \partial _\mu \bar \phi \partial ^\mu \bar \phi ^\dagger +
{i \over 2 }  ( \bar \phi ^\dagger v \cdot \partial  \bar \phi -
\bar \phi v \cdot \partial  \bar \phi ^\dagger )  +
{\lambda ^2 \over 2 } \tilde \pi ^4 - 2{ \lambda ^2 \over M}
\tilde \pi ^2 \bar \phi \bar \phi ^\dagger \nonumber \\
&+& 2 i { \lambda ^2 \over M ^2 } \left[
\bar \phi ^\dagger \tilde \pi v \cdot \partial 
(\tilde \pi \bar \phi)  -
\bar \phi \tilde \pi v \cdot \partial (\tilde \pi \bar \phi ^\dagger)
\right]  +
 { \lambda ^2 \over M^2 }
\bigg( -{1\over 2 }\tilde \pi ^2 \Box \tilde \pi ^2 +
{1\over M} \bigg\{4
 \lambda ^2 \tilde \pi ^4 \bar \phi  \bar \phi ^\dagger \nonumber \\
 &-&
( \tilde \pi \bar \phi \Box (\tilde \pi \bar \phi ^\dagger) +
\tilde \pi \bar \phi ^\dagger  \Box (\tilde \pi \bar \phi) ) +
4[ \tilde \pi \bar \phi (v\cdot \partial)^2 
(\tilde \pi \bar \phi ^\dagger) +
\tilde \pi \bar \phi ^\dagger  (v\cdot \partial)^2
(\tilde \pi \bar \phi) ] 
\bigg\} \bigg) \label{eflagM2}
\end{eqnarray}
In (\ref{eflagM2}) we have kept terms which apparently are of
order $ {\cal O} (1/ M^3)$. The reason is that for Green functions
involving $\bar \phi \bar \phi^\dagger$, an extra factor $2M$ has to
be introduced, (\ref{nrPHI}).
Clearly, we have generated HMT vertices for the Green functions
discussed at the beginning of this section. Two four--point vertices are not
suppressed in the limit $M \rightarrow \infty$. In fact, they are the
only non-suppressed Green functions in this limit.
The four pion function has been obtained by
integrating out the {\em low} component, $\tilde \psi$,
of the {\em heavy} particle, $\phi$. This
is the same attitude followed for example in \cite{TONInucl},
where the vector mesons where integrated out to obtain their contributions
to the pseudo-scalar interactions at low-energy. The other four--point
Green function, $\tilde \phi\tilde\phi^\dagger\pi\pi$,  has been obtained
by integrating out the {\em high}
components, $ \tilde \sigma $ and $ \tilde \sigma ^\dagger $ of the
{\em light} particle, the $\pi$. To our knowledge, this procedure was
used for the first time in\cite{BGT}, there we used a diagrammatic approach.

We also find a six point function.
It is suppressed by a factor
$1/M^2$. 
%If we continue our expansions of $\tilde \psi$,
%$\tilde \sigma$, (\ref{sol1OM}) and ${\cal L}_1^E$, (\ref{eflagM2})
%to higher orders in ${\cal O}(1/M)$, we would find eight, ten,
%and higher order point functions.
We have checked that the four and six point functions
generated by (\ref{eflagM2}) reproduce exactly, up to order 
${\cal O} (1/M^2)$
the corresponding Green functions of the full relativistic theory,
(\ref{model}). 

Let us have a closer look at the situation
of the six point functions in general, see Fig.~\ref{figgreen}. The
first three diagrams, $a,b$ and $c$ are diagrams in the full theory,
(\ref{model}), while diagram $d,e$ and $f$ live in the effective theory.

Cutting the internal heavy meson of diagram $a$, we obtain two
four point functions that lie within the HMT.
% As we have just seen,
The HMT can describe these two four--point functions, (\ref{eflagM2}).
The HMT 
counterpart for diagram $a$ is diagram $d$. The situation is very
similar for diagram $b$: here, we again obtain two four point functions
that
are described within the HMT by cutting one of the internal pion line.
The fact that one of these is a four pion function shows why at the
beginning of the section we claimed the processes
$n\,\pi \longrightarrow \,k\,\pi $ also have to be taken into account.
 The HMT counterpart of $b$ is $e$. The situation
is however different for diagram $c$. Here no HMT diagram can be
generated by cutting any of the internal lines. In the HMT, we can not
construct its counterpart from ``smaller'' diagrams. A new vertex, the
six point vertex in (\ref{eflagM2}) has been generated to account for
it. The HMT counterpart for $c$ is $f$.

 This can be generalized, and if we continue our expansions of
$\tilde \psi$,
$\tilde \sigma$, (\ref{sol1OM}) and ${\cal L}_1^E$, (\ref{eflagM2})
to higher orders in ${\cal O}(1/M)$, we would generate new
eight, ten, and higher order vertices.

\section{One loop matching in the Simple Model}

We will now illustrate with a few examples that the effective 
lagrangian we have constructed in the previous section, which does
reproduce correctly the tree level Green functions of the full
relativistic theory, also reproduces correctly the non-analytic
parts of the Green functions at the one loop level. 
The first non-trivial results in that model show up in four--point
functions. We add a pion mass\footnote{For the quantities and to the order
considered here the HMT Lagrangian is identical to the one of
Sect. \protect{\ref{treesimple}}.}
 and a coupling of pions to an
external scalar field $S(x)$ to the model. Then we have nontrivial
behaviour already in two-- and three--point functions.
The Lagrangian becomes
\begin{equation}
{\cal L} = {\cal L}_0+
{1\over 2} m^2 \pi ^2 -
S(x) \pi ^2\,.
\label{lagSOURCE}
\end{equation}

The first
example will be the one--loop self-energy of the $\phi$, at leading
order in $1/M$. In the effective theory (HMT), its contribution is
given by
\begin{equation}
\Pi_{\phi}^{E} =
-4 i \lambda ^2 {m^2 \over 16 \pi ^2}
\left( -{1\over \epsilon} +\gamma _e 
-\log 4\pi -1  + \log {m^2 \over \mu ^2 } \right) + {\cal O}(1/M^2)
\label{phiselfEFF}
\end{equation}
Where we have followed the standard notation and used
dimensional regularization with $D=4-2\epsilon$
and $\gamma_e$ is Euler's constant.

It is also easy to obtain the result in the full relativistic theory:
\begin{eqnarray}
\Pi_{\phi}^{F} &=& 2 \lambda ^2 M^2 {i \over 16 \pi ^2}
\left( {1\over \epsilon} + \log 4 \pi -\gamma _e - 
\log { M^2 \over \mu ^2 } \right) \nonumber \\
& + & 4 \lambda ^2 m^2 {i \over 16 \pi ^2} \left(
1 -\log {-m^2 \over \mu ^2 } + \log { M^2 \over \mu ^2 } \right) +
{\cal O}(1/M^2)
\label{phiselfFULL}
\end{eqnarray}
Let us make a few observations here:
\begin{enumerate}
\item
The nonanalytic terms we are interested in are the logarithms of
$m^2$. It follows from (\ref{phiselfEFF}) and (\ref{phiselfFULL}) that
the effective theory does reproduce this. This is the main point.
\item
The analytic dependence on $m^2$ gets a correction at one--loop compared
to the tree--level one. This correction is even infinite in the $M\to\infty$
limit. This dependence cannot be obtained from the HMT. We have to leave
a term proportional to $m^2$ for the $\phi$ mass in the HMT.
\item
It is interesting to notice that in the full theory, 
$ \Pi_{\phi}^F $ has an imaginary part, which is absent in the
effective theory. This is due to the fact that in the full theory
the $\phi$ has a finite width due to the decay $\phi \rightarrow 2\pi$.
Since this decay cannot be described within the heavy meson theory,
no imaginary part is present in $\Pi_{\phi}^{E}$. 
A small discussion about this fact can be found in Sect. \ref{width}.
\end{enumerate}

\begin{figure}
\begin{center}
\leavevmode\epsfxsize=12cm\epsfbox{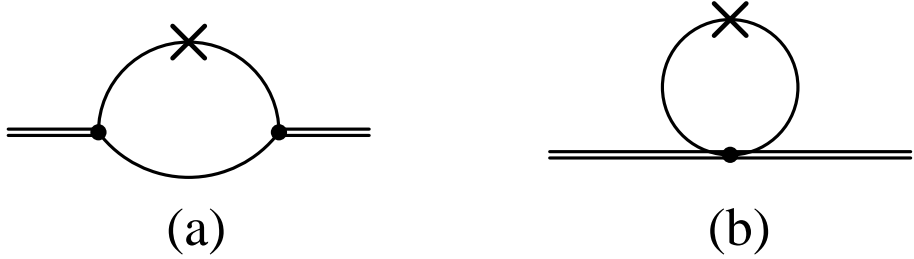}
\end{center}
\caption{\label{figtoy} Scalar form-factor in the Simple Model: (a)
 relativistic theory (b) effective theory. Double lines correspond to
the $\phi$ field, full lines to the $\pi$ field, and crosses to insertions
of the scalar source $S(x)$.} 
\end{figure}

As a second example we consider the scalar form--factor of the $\phi$,
with the scalar source $S(x)$ defined above.
To leading order in $1 / M$, the only diagram that contributes in
the effective theory is the one of Fig.~\ref{figtoy}b.  It contributes 
\begin{equation}
{ 8 i \lambda ^2 \over (4 \pi)^{2} }
\left[\frac{1}{\epsilon}-\gamma_e+\log(4\pi)
-\int _0 ^1\log \left({m^2-q^2 x(1-x)-i\varepsilon \over \mu ^2} 
\right)dx\right]
\label{efmatching}
\end{equation}
where $q$ is the momentum that flows through the scalar source. 

In the
full theory, we have to consider only the diagram of Fig. \ref{figtoy}a
at this order.
Its complete
contribution reads
\begin{equation}
I = {8i \lambda ^2 M^2 \over (4 \pi)^2 } 
\int _0 ^1 
{ y dx dy \over 
( -m^2 +y(1-y)[Q^2 -q^2x] +[xy - (xy)^2 ] q^2+i\varepsilon ) }
\label{fullMATCH1}
\end{equation}
As in the previous case, $q$ is the momentum that enters through the
source $S(x)$, and $Q$ is the momentum of the heavy field, $\phi$, 
$Q^2 = (Q-q)^2 = M^2$. 
We are
only interested in the leading order in $1/M$. The contributions
away from $y\approx 0$ and $y\approx 1$ can always be expanded in $1/M^2$,
while we expand in $(1-y)$ and $y$ near $1$ and $0$ respectively.
\begin{eqnarray}
I&\approx&
{8i \lambda ^2 M^2 \over (4 \pi)^{2} }
\int _0 ^1 dx \Bigg\{ \int _{1 -\delta} ^1 
{  dy \over 
( -m^2 +(1-y)Q^2 +x( 1- x ) q^2 +i\epsilon) }
\nonumber\\&&
+\int^{1-\delta}_\alpha\frac{ydy}{y(1-y)Q^2}
+\int_0^\alpha\frac{ydy}{-m^2+yQ^2+i\varepsilon}
\Bigg\}\nonumber\\
&=&\! {8i \lambda ^2 M^2 \over (4 \pi)^{2} }
\int_0^1 dx \left(-\frac{1}{M^2}\log\left|\frac{m^2-x(1-x)q^2}{M^2}\right|
 - i\,\pi\int_0^\delta dz\,\delta(m^2-x(1-x)q^2-z Q^2)\right)
\nonumber\\ 
\label{fullMATCH2}
\end{eqnarray}
where $\delta$ and $\alpha$ are small numbers, but much larger than 
$\max(m^2,|q^2|)/M^2$.
After integrating over $y$ we can again make the same three remarks made above:
\begin{enumerate}
\item
The nonanalytic term in $m$ and $q$ are  
exactly the same as in the integral of (\ref{efmatching}).
\item
The argument of the logarithm is $M^2$. To obtain $\mu^2$ we
have to add a term proportional to 
$\log(M^2/\mu^2)$. This term requires us to introduce a direct 
$S(x)\tilde\phi\tilde\phi^\dagger$ coupling in the HMT to obtain matching.
\item
The imaginary part only coincides provided the part corresponding to
the $\phi$ width can be neglected. It is equal to
$- \lambda ^2 M^2 / (2 \pi)\left\{  \theta (1- 4 m^2 / q^2)
\sqrt{1- 4 m^2 / q^2}-1\right\}$.
The second term is the imaginary part due to the $\phi$-width and the
first part is the long--distance part present in the effective theory
(\ref{efmatching}).
\end{enumerate}

\section{A Chirally Symmetric Model}
\label{chiralmodel}

The QCD lagrangian has an approximate
chiral symmetry $SU(N_f)_L\times SU(N_f)_R$ for
$N_f$ light flavours. This implies the existence of light pseudo-goldstone
Bosons, which we will refer to as pions. The method of dealing with these
at low energy 
is Chiral Perturbation Theory\cite{GL3}. In (\ref{model}) there
were only vertices with a limited number of legs. Effective lagrangians
with pions tend to have vertices with any number of legs. The method
illustrated on the simple example becomes much more powerful here.
In this section we will introduce a pion lagrangian coupling to an external
scalar source $S(x)$ and we will show how to obtain an effective lagrangian
with terms bilinear in $S(x)$ to reproduce all Green functions of the type
$S(-p)S(p+k)\pi^n$ with $p$ a large momentum and $k$ and the $\pi$-momenta
small. Again we will restrict our discussion to tree level.

Consider the following lagrangian: 
\begin{equation}
{\cal L}_2 = \frac{F^2}{4}\langle u_\mu u^\mu + S(x)\chi_+\rangle\,.
\label{defL2}\end{equation}
Here we used the following notation
\begin{eqnarray}
\chi_{\pm}&=& u^\dagger_R\chi u_L \pm u_L^\dagger \chi^\dagger u_R
\nonumber\\
u_\mu &=& i
\left( u_R^\dagger(\partial _\mu  -i r_\mu)u_R
-u_L^\dagger(\partial _\mu -il_\mu)u_L\right)
\nonumber\\
\Gamma_\mu &=& \frac{1}{2}
 \left( u_R^\dagger(\partial _\mu -i r_\mu)u_R
+u_L^\dagger(\partial _\mu -il_\mu)u_L\right)\,
\label{definitions}
\end{eqnarray}
and $\langle {\cal C} \rangle$ denotes the trace of ${\cal C}$.
The external fields $\chi, l_\mu,r_\mu$ are defined in the usual way,
see \cite{TONInucl}. 
$\chi$
contains the quark masses. All of these are $N_f$ by $N_f$ matrices
in flavour space. The matrices $u_R$ and $u_L$ are given in terms of the pions
as 
\begin{equation}
u_R = u_L^\dagger = u = \exp\left(\frac{i}{F}\pi^a\lambda^a\right)\,.
\label{defu}
\end{equation}
The transformations under a chiral transformation $g_L\times g_R\in 
SU(N_f)_L\times SU(N_f)_R$ are given by
\begin{equation}
u_R\to g_R u_R h_c^\dagger,\quad u_L\to g_L u_L h_c^\dagger,\quad
u_\mu\to h_c u_\mu h_c^\dagger\quad\mbox{and}\quad
S(x) \to h_c S(x)  h_c^\dagger\,.
\label{transformations}
\end{equation}
This together with (\ref{defu}) determines $h_c$ in terms of the pion fields
$\pi^a$ and $g_L,g_R$.

We now set
\begin{equation}
S(x) = e^{-i Mv\cdot x}\tilde S(x) + e^{i M v\cdot x}\tilde S(x)^\dagger
\label{defSt}
\end{equation}
where $\tilde S(x)$ only has momenta small compared to $Mv$ and look at
Green functions with one insertion of $\tilde S(x)$ and one of
$\tilde S(x)^\dagger$. These have contributions with pions of momenta of order
$Mv$. We treat these pions in the same way as in the previous section
by introducing a spurious field that takes its role.
The difficulty
here is to do it in a manifestly chirally invariant fashion.

This can be done by introducing a ``hidden'' $SU(N_f)_3$ symmetry that
is rather nonlinearly realized.
 We first introduce three $N_f$--by--$N_f$ special
unitary matrices, $u_{\cal R}$, $u_{\cal L}$ and
$W=w^2$ transforming under $(g_L\times g_{R}
 \times h_1\times h_2)\in 
SU(N_f)_{L}\times SU(N_f)_{R}\times SU(N_f)_1\times
SU(N_f)_2$ as:
\begin{equation}
u_{\cal R}\to g_{ R}  u_{\cal R} h_1^\dagger,\quad 
u_{\cal L}\to g_{ L} u_{\cal L} h_2^\dagger,\quad
W\to h_1 W h_2^\dagger
\end{equation}
and we define $h_3$ via
\begin{equation}
w = \sqrt{W}\to h_1 w h_3^\dagger = h_3 w h_2^\dagger\,.
\end{equation}
The other fields are defined via $u_R =  u_{\cal R} w$ and 
$u_L = u_{\cal L} w^\dagger$
and transform as in (\ref{transformations}) with $h_c$ replaced by $h_3$.
We remove one of the spurious degrees of freedom by fixing $h_1$ such that
$u_{\cal R} = u_{\cal L}^\dagger$, which is always 
possible. The extra symmetry is then the
$h_2$ one with $h_1$ fixed in this way.
The remaining extra degree of freedom is of course also spurious since it can
be removed by the choice of $h_2$.
So we have
\begin{equation}
u_R = u w\quad\mbox{and}\quad u_L = u^\dagger w^\dagger\,.
\label{defu2}
\end{equation}
Notice that with this $w$ transforms as a pseudoscalar
and plays the same role as $\sigma$ in  section \ref{treesimple},
and $u_R$ and $u_L$ take the role of $\Pi$.
We should now use the quantities in (\ref{defL2}) and (\ref{definitions})
with (\ref{defu2}) instead of (\ref{defu}).
We also use the quantities defined in Eq. (\ref{definitions}) with
a superscript $l$ to indicate those quantities with $w=1$ (i.e. the 
{\em low} components). For the processes
we are interested in we expand $w = \exp(i \xi/2)$ up to
second order in $\xi$, there never are more than two lines with a high momentum
at any vertex in any diagram.

The equation of motion derived from ${\cal L}_2$ is
\begin{equation}
\nabla_\mu u^\mu = \frac{i}{4}\left(\{S(x),\chi_-\}-\frac{2}{N_f}\langle S(x)
\chi_-
\rangle\right)\,,
\label{EOM1}
\end{equation}
or expanded to second order in $\xi$:  
\begin{eqnarray}
&&
\nabla_\mu^l u^{l\mu} - (\nabla_\mu^l)^2\xi
-\frac{1}{4}[u^l_\mu,[u^{l\mu},\xi]]
+\frac{1}{2}[\nabla^l_\mu \xi,[u^{l\mu},\xi]]
-\frac{1}{8}[\xi,[\xi,\nabla^l_\mu u^{l\mu}]]  =  \nonumber\\ 
&&\frac{i}{4}\left(\{S(x),\chi_-^l\}-\frac{2}{N_f}
\langle S(x)\chi_-^l\rangle\right)
+\frac{1}{8}\left(\{S(x),\{\xi,\chi_+^l\}\}
  -\frac{2}{N_f}\langle S(x)\{\xi,\chi_+^l\}\rangle\right)
\nonumber\\
&&+{\cal O}\left(\xi^3,S(x)\xi^2\right)\,.
\label{EOM2}
\end{eqnarray}
Here we have defined $\nabla^l_\mu C   = \partial_\mu C + [\Gamma^l_\mu,C]$.

In order to allow a consistent approximation of the quantities only 
containing $u,\chi,l_\mu,r_\mu$ we now choose the ``gauge'' for $w$ or 
$\xi$ such that
\begin{equation}
(\nabla_\mu^l)^2\xi +\frac{1}{4}[u^l_\mu,[u^{l\mu},\xi]]
 =  
\frac{-i}{4}\left(\{S(x),\chi_-^l\}-\frac{2}{N_f}\langle S(x)
\chi_-^l\rangle\right)
+{\cal O}\left(\xi^3,S(x)\xi^2\right)\,.
\label{gauge}
\end{equation}
We can now see that if we restrict $S(x)$ as in Eq. (\ref{defSt}) 
and only consider
the Green functions mentioned above we obtain a {\em consistent} approximation
scheme using
\begin{equation} 
\label{pseudoapprox}
u=\tilde u\quad{\rm and}\quad
\xi=e^{-iMv\cdot x}\tilde\xi + e^{iMv\cdot x}\tilde \xi^\dagger\,
\end{equation}
with the tilded quantities with low momenta,
since we can use our choice of gauge (\ref{gauge}) to remove the terms
out of the equation of motion for the field $u$ that would have required
a high momentum.
The equation for $\tilde\xi$  starts only at the level $1/M^2$ - this
fact will simplify extremely the lagrangian - and can be
solved iteratively. Its first two terms are given by:
\begin{equation}
\tilde \xi = \frac{1}{M^2}  \left(\ \frac{i}{4}
(\{ \tilde S(x),\chi_-^l \}-\frac{2}{N_f} \langle\tilde S(x) \chi_-^l \rangle)
+\frac{1}{2M} \nabla_\mu^l (v^\mu
(\{ \tilde S(x),\chi_-^l \}-\frac{2}{N_f} 
\langle\tilde S(x) \chi_-^l \rangle)) 
\right)\
\label{scalarxi}
\end{equation}
And the resulting effective lagrangian to ${\cal O}(1/M^2)$
is
\begin{equation}
{\cal L}_2^E = \frac{F^2}{4} \left[\langle u_\mu^l u^{l\mu}\rangle
+\frac{1}{8 M^2} \left(\langle 
\{\tilde S(x),\chi_-^l\} \{\tilde S(x)^\dagger,\chi_-^l\}\rangle
-\frac{1}{N_f} \langle \{\tilde S(x),\chi_-^l\} \rangle 
\langle \{\tilde S(x)^\dagger,\chi_-^l\} \rangle\right)\right]
\end{equation}

The advantage of splitting the fields in such a way
is that chiral symmetry remained
manifest throughout the whole calculation.
  
\section{Vector Mesons}

We will now apply our method to the vector meson case. Here the 
situation is a little bit more involved because of the structure of the
heavy meson propagator in the nonrelativistic limit, and we will
construct the effective theory in two steps: First, we integrate out
the non-relevant degrees of freedom, that is, the low component of the
vector mesons and the high components of the pseudoscalars. Then,
project the theory on the ``orthogonal subspace'' (see section IV
of \cite{BGT}), extract the large momentum proportional to $M_V$ from
the vectors and construct an effective lagrangian that allows to perform
a consistent expansion in $1/M_V$
We again
restrict ourselves to the processes of the type:
\begin{equation}
\label{vecprocesses}
\mbox{Vector }+\, n\,\pi \quad\longrightarrow\quad\mbox{Vector }+\,k\,\pi\,.
\end{equation} 
There exist several models that describe the interaction of vector mesons
with the pseudoscalar Goldstone bosons, see \cite{TONI}
and references therein. We will present
our result in terms of model III of that reference (see below
for a short discussion about other models).\\
Then the lagrangian
reads:
\begin{equation}
\label{vecL}
{\cal L}_3 = \frac{F^2}{4}\langle u_\mu u^\mu \rangle - \frac{1}{4} 
\langle \overline V_{\mu\nu} \overline V^{\mu\nu} \rangle
+\frac{M^2}{2} \langle(\overline V_\mu-\frac{i}{g} \Gamma_\mu)^2\rangle
\end{equation}
With $\overline{V}_{\mu\nu}=\partial_\mu\overline{V}
-\partial_\nu\overline{V}_\mu
-ig\left[\overline{V}_\mu\,,\,\overline{V}_\nu\right]$.
We have disregarded the terms containing quark masses for the present
discussion, because they are ${\cal O}(1/M)$ suppressed. 
The pseudoscalar fields transform as in (\ref{defu}),
and the vector fields transform as:
\begin{equation}
\label{vectransf}
\overline V_\mu  \to h_c \overline V_\mu h_c^\dagger 
+\frac{i}{g} h_c \partial_\mu h_c^\dagger\,.
\end{equation}
where $h_c$ is defined in (\ref{transformations}).

We now enlarge the symmetry for the pseudoscalar sector in the same way
as we did  
in the previous section
to obtain an extra degree of freedom,
$u_R=u_L^\dagger= u w$, and a ${\cal R}^4$
hidden symmetry to introduce a spurious vector 
degree of freedom (corresponding to
the $\psi$ field in Sect. \ref{treesimple}). We have as a total set of
transformations:
\begin{eqnarray}
\alpha_\mu\in{\cal R}^4 &:& W_\mu\to W_\mu+\alpha_\mu\nonumber\\
&& X_\mu \to X_\mu-\alpha_\mu\\
h_3&:& W_\mu \to h_3 W_\mu h_3^\dagger\nonumber\\
&& X_\mu \to h_3 X_\mu h_3^\dagger +\frac{i}{g}h_3\partial_\mu h_3^\dagger\,.
\label{vectransf2}
\end{eqnarray}
We now use $\overline{V}_\mu = W_\mu + X_\mu$ together with (\ref{definitions})
and (\ref{defu2}) in (\ref{vecL}), with $h_3$ defined as in
Sect. \ref{chiralmodel}.
The model is then exactly equivalent to the original model (\ref{vecL})
setting $X_\mu=0$ and $\xi=0$.

Notice that the non-linear term in (\ref{vectransf2})
has been chosen inside $X_\mu$, the reason is that later 
we will choose the ${\cal R}^4$
gauge such that this field becomes the ``low'' momentum component of
the vector. It therefore has to take the low momentum nonlinear term.
In
principle, for the vectors we would have to use the general parametrization
(\ref{simple}), and we would have to integrate its low component out.

The equation of motion for the pseudoscalar fields reads:
\begin{equation}
\label{eompseudo}
\nabla_\mu u^\mu -\frac{iM^2}{2gF^2} [\overline V_\mu-\frac{i}{g} \Gamma_\mu,
u^\mu] = 0\,.
\end{equation}
and the vector equation of motion:
\begin{equation}
\partial^\mu \overline{V}_{\mu\nu}
-ig[\overline V^\mu,\overline V_{\mu\nu}]+M^2(\overline V_\nu-\frac{i}{g}
\Gamma_\nu) = 0\,.
\end{equation}
We now expand these equations up to second order in the $W$
and $\xi$ fields\footnote{This is all we need for the processes 
(\protect{\ref{vecprocesses}}) at tree level, when later $W$ and $\xi$
become the ``high'' momentum components.}. Choosing the gauge fixing
conditions such that for the processes (\ref{vecprocesses}) $X_\mu$ only
has ``low'' momentum and $\xi$ only ``high'' momentum:
\begin{eqnarray}
&&-(\nabla^l_\mu)^2 \xi-\frac{1}{4} [u^l_\mu,[u^{l\mu},\xi]]
\nonumber\\&&
-\frac{iM^2}{2gF^2} \left(\left[W_\mu+\frac{i}{4g}[\xi,u^l_\mu],u^{l\mu}\right]
-\left[X_\mu-\frac{i}{g}\Gamma^l_\mu\,,\,\nabla^{l\mu}\xi\right]\right)
+ {\cal O}((\xi,W)^3)=0\,,
\label{gaugepseudo}
\end{eqnarray}
for the pseudoscalars and
\begin{eqnarray}
&&
\nabla^{X\mu} X_{\mu\nu}-ig\nabla^{X\mu}[W_\mu,W_\nu]-ig[W_\mu,
\nabla^{X\mu} W_\nu-\nabla^{X}_{\nu} W^\mu]\nonumber\\&&
+M^2\left(X_\nu-\frac{i}{g}\Gamma^l_\nu
-\frac{i}{8g}[\xi,\nabla^l_\nu\xi]\right)
+ {\cal O}((\xi,W)^4)=0\,,
\label{gaugevector}\end{eqnarray}
for the vectors. The superscripts $l$ denotes quantities with $w=1$ as
in Sect. \ref{chiralmodel} and
we have defined the following covariant derivative:
\begin{equation}
\nabla^X_\mu {\cal A} = \partial_\mu {\cal A} -ig [X_\mu,{\cal A}]\,.
\end{equation}

This choice now allows us to make the {\em consistent} set of
approximations
\begin{equation}
\sqrt{2M}\; W_\mu = 
e^{-iMv\cdot x} \tilde W_\mu + e^{iMv\cdot x} \tilde W_\mu^\dagger
\quad\mbox{and}\quad
X_\mu=\tilde X_\mu
\label{effectW}
\end{equation}
together with those of (\ref{pseudoapprox}).
Again, tilded symbols means that they are restricted to low momenta.

Solving iteratively the gauge conditions we find the following
solutions for the spurious fields, using $g = M/(2F)$ \cite{TONI}
to count $g$ as ${\cal O}(M)$:
\begin{eqnarray}
&& \xi = \frac{i}{2gF^2} [ W_\mu,u^{l\mu}]+\ldots\
\nonumber\\&&
\tilde X_\mu = \frac{i}{g}\Gamma_\mu^l+\ldots\,.
\end{eqnarray}
Here we only keep the leading 
terms contributing to (\ref{vecprocesses}), and set the external 
sources $l_\mu$ and $r_\mu$ to zero.

Putting these solutions inside the lagrangian (\ref{vecL}),
after some algebra, we obtain
\begin{eqnarray}
{\cal L}_3^E&=& 
-{1\over 2} \langle\nabla_\mu^l W_\nu \nabla^{l\mu} W^\nu\rangle
+{1\over 2}\langle\nabla^l_\mu W^\mu \nabla^l_\nu W^\nu\rangle
+\frac{F^2}{4}\langle u^l_\mu u^{l\mu}\rangle
+\frac{1}{4}\langle [W_\mu,u^{l \mu}] ^2 \rangle \nonumber\\&&-
\frac{1}{4}\langle [u^l_\mu,u^l_\nu][W^\mu, W^\nu]  \rangle
 + {M^2 \over 2} \langle W_\mu W^\mu  \rangle +
\ldots .
\label{veceff}
\end{eqnarray}
The dots in (\ref{veceff}) denote terms with zero or more than two
vectors, as well as terms suppressed in the $1/M$ counting. The term
$(1/4) \langle [ W_\mu u^{l \mu} ] ^2 \rangle $ is generated by the
high component of the pions, $\xi$, while the the rest of the terms
(excluding the mass term) are generated by the low component of the
vectors. 

This last step to be taken is to go from (\ref{veceff}) to an
effective theory which allows us to perform a $1/M$ expansion. The
presence of terms involving $\partial _ \mu W ^\mu$ in (\ref{veceff})
makes this last step a little bit more subtle, and we cannot simply
replace (\ref{effectW}) in (\ref{veceff}). We will follow the method
of section 4 of \cite{BGT} and introduce a parallel component,
$ \tilde W_\parallel^\mu=v^\mu(v\cdot \tilde W)$,
and a perpendicular component 
$ \tilde W_\perp^\mu=\tilde W^\mu-v^\mu (v\cdot \tilde W)$,
and integrate the parallel component out,
see \cite{BGT} for details. To leading order in $1/M$, and in 
the notation of \cite{BGT} (Eq.(13) of that paper), this leads to
\begin{equation}
a_2 = -\frac{1}{2M}, \quad a_3 = 0, \quad a_7 =-\frac{1}{4M} ,
 \quad a_8 = 0,
\quad a_9 = \frac{1}{4M}
\label{newAS}
\end{equation}
We have checked that these terms can be reproduced
{\em exactly} diagrammatically.
The fact that $a_3$ vanishes in (\ref{newAS}) might seem surprising,
since in (\ref{veceff}) the term 
\begin{equation}
\frac{1}{2} \langle \nabla^l_\mu W^\mu  \nabla^l_\nu W^\nu\rangle
\label{strangeA3}
\end{equation}
is present. To understand $a_3 = 0$, we first notice that
(\ref{strangeA3})
does not contribute to pion-vector scattering, since for on-shell vectors,
$\partial _ \mu W^\mu = 0$. For processes such as
$2 \pi V \rightarrow 2 \pi V$, the contribution
of (\ref{strangeA3}) to the diagram of Fig. \ref{figgreen}f does not vanish.
To leading order in the $1/M$ expansion, this is cancelled by the
contribution of (\ref{strangeA3}) to the diagram of Fig. \ref{figgreen}d.
Contributions where only one of the vertices of Fig. \ref{figgreen}d is
generated by (\ref{strangeA3}) are suppressed in  $1/M$. We conclude
that (\ref{strangeA3}) does not contribute to the process 
$2 \pi V \rightarrow 2 \pi V$ at leading order in $1/M$, and it can be
seen that this holds for  $n \pi V \rightarrow m \pi V$ for
any $n$, $m$. Eq. (\ref{newAS}) agrees with \cite{BGT} after correcting the
misprints there.

In this section we have used model III of \cite{TONI}, but similar
methods can be applied to the other models. We have checked that all
three models predict the same $\langle [ W_\mu , u ^\mu ] \rangle$ term,
(coming from the integration out of the high component of the pions)
which contributes to the masses in \cite{BGT},
but different 
$\langle [u_\mu , u_\nu][ W^\mu , W ^\nu ] \rangle$ terms, 
(integration out of the low components of the vectors) which do
not contribute to the masses in \cite{BGT}. The 
coefficient we find for this term in model III is $1/4$, while
in model II we find $1/8$. The fact that different models predict
different ${\cal O} (V^2)$ lagrangians should not be surprising, since
${\cal L}_{II}$ and ${\cal L}_{III}$ in \cite{TONI}  differ by terms
of order ${\cal O} (V^2)$.

\section{Vectors at one loop}

Contributions to the vector masses in the relativistic and the heavy meson
formulation are given by the diagrams in Fig. \ref{figtoy} with the crosses
removed. We introduce here a pion mass term, $m=m_\pi$,
to have relevant nonanalytic
contributions and only consider the pion contributions.
In the effective formulation we obtain a contribution to the $\rho$ mass-shift
\begin{equation}
\delta M = -\frac{m^4}{32\pi^2 M F^2}
\log\left(\frac{m^2}{\mu^2}\right)\,.
\end{equation}
In the relativistic formulation we obtain
\begin{equation}
\delta(M^2) =
\frac{M^4}{2 g^2 F^4}B_{20}(M^2,m^2)\,.
\end{equation}
with
\begin{equation}
\int\frac{d^dp}{i(2\pi)^d} \frac{p_\mu p_\nu}{(p^2-m^2)((p+Q)^2-m^2)}
= g_{\mu\nu}B_{20}(Q^2,m^2)+Q_\mu Q_\nu B_{22}(Q^2,m^2)\,.
\end{equation}
We can now check by expanding $B_{20}(M^2,m^2)$ that the first nonanalytic
dependence on $m^2$ only appears at order $m^4/M^2\,\log(m^2/M^2)$
and that the coefficients agree if we use the value of $g$ used above.
So here we have an indication that the procedure of the heavy meson
theory also works in this case at one loop.

\section{Width}
\label{width}

The width of the ``heavy'' particle due to its decays is not
included in the HMT. Here we discuss shortly under what circumstances
we expect the HMT to be useful given that it cannot easily describe the width.
In the tree level processes in the relativistic theory we can describe
the width of the heavy particle by using as propagator instead
\begin{equation}
\frac{1}{p^2-M^2+i M\Gamma}\approx \frac{1}{M}\,\frac{1}{2 v\cdot k+i\Gamma}
\approx \frac{1}{2M}\frac{1}{v\cdot k}
\left\{1-\frac{i\Gamma}{2v\cdot k}+\cdots\right\}\,.
\label{propagator}
\end{equation}
Here we see that if the typical off-shellness of the ``heavy'' particle
is large compared to its width the latter can be neglected.
For vector mesons at tree--level this will always be the case except
for the $\rho$. But even there we are helped by the extra factor
of two in the expansion in (\ref{propagator}).

In the loop diagrams a similar argument will hold if the contributions
of the integrals very near to the mass-shell is small compared to the others.
Again for vector mesons we expect this to be the case except possibly
for the $\rho$. Even for the $\rho$, the dependence on the strange quark mass
and similar effects come from intermediate states that are far off-shell so
those should be reliably estimated in the HMT.

\section{Conclusions}

The problem of reducing a relativistic theory to a ``heavy'' effective
formulation restricted to a particular type of processes has been solved
in the case where the number of ``heavy'' particles is not conserved.
The problem of ``high'' components of light particles and ``low'' components
of heavy particles has been treated in a natural way. We have enlarged the
symmetry by a ``hidden'' symmetry. The spurious degrees of freedom thus
introduced, can be chosen by a particular choice of ``gauge'' for the
extra symmetry to play the role of the ``high'' components of the light
particle and of the ``low'' components of the heavy particle.

The choice of gauge allows then for a simple reduction to the
``heavy'' effective theory. 
We explicitly matched all relevant Green
functions at tree level, first in a toy model and afterwards
in two Chiral models,
showing that matching already at this level is rather subtle. In the
last two examples our method allows
to have chiral symmetry explicit during all stages of the calculation.
The calculation for the model for the vector mesons agrees with
our previous matching procedure\cite{BGT} which was done by
matching specific Green functions.

We have shown that with this procedure the non-analytical parts at one-loop
level are also recovered in a few examples.
We included both a two-- and a three--point function in the simple model
and a two--point function in the vector chiral case.
These examples provide support that the ``heavy'' meson theory will
also be correct at the quantum level.

We discussed shortly the relevance of the nonzero
width of the ``heavy'' particle.

\section*{Acknowledgments}
PG acknowledges a grant form the Spanish Ministry for
Education and Culture. \\
PG and PT thank the hospitality of the Theoretical Physics 
Department of Lund University, where part of
this work has been carried out.

\end{document}